\documentstyle[prl,aps,epsf]{revtex}

\newcommand{\be}{\begin{equation}}
\newcommand{\ee}{\end{equation}}
\newcommand{\bea}{\begin{eqnarray}}
\newcommand{\eea}{\end{eqnarray}}

\newcommand{\rf}[1]{(\ref{#1})}

\begin{document}
\draft
\title{Generalized Second Law of Black Hole Thermodynamics \\
and Quantum Information Theory}
\author{A. Hosoya, A. Carlini and T. Shimomura}
\address{Department of Physics,
Tokyo Institute of Technology,
Oh-Okayama, Meguro-ku, Tokyo 152-8550, Japan}
    \twocolumn[\hsize\textwidth\columnwidth\hsize\csname
    @twocolumnfalse\endcsname
\maketitle
\begin{abstract}
  We propose a quantum version of a gedanken experiment which supports
the generalized second law of black hole thermodynamics.
A quantum measurement
of particles in the region outside of the event horizon decreases the
entropy of the outside matter due to the entanglement of the
inside and outside particle states.
This decrease is compensated, however, by the increase in the detector entropy.
If the detector is conditionally dropped into the black hole depending on the
experimental outcome,
the decrease of the matter
entropy is more than compensated by the increase of the black hole
entropy via the increase of the black hole mass which is ultimately
attributed to the work done by the measurement.
\end{abstract}


\pacs{PACS numbers: 03.67.-a, 04.70.Dy, 04.70.-s, 05.70.-a}
    \vskip 3ex ]

\narrowtext


\section{Introduction}

The striking parallelism \cite{BCH,BE1,HW,W} between black hole physics and
the laws of thermodynamics has attracted many physicists. However, it
seems there remains some controversy concerning the question
to what extent the paralellism works for the second law.
The generalized second law of black hole thermodynamics states that the
sum of the Bekenstein-Hawking black hole entropy, which is a quarter
of the area $A$ of the event horizon, and the ordinary thermodynamical
entropy $S^m$, i.e.
\begin{eqnarray}
S^{tot}={A\over 4}+S^m~,
\label{eqn:gsl}
\end{eqnarray}
never decreases.

One of the directions in the study of the generalized second law is based
 on a gedanken experiment suggested by Bekenstein \cite{BE1,BE2} and
discussed by many people.
A box of mass $E^{b}$ and entropy $S^{b}$ is lowered by a string from
infinity to a point near the horizon.
The box is then thrown into the black hole.
Apparently, the entropy contained in the box is gone for the
outside region of the black hole.
This decrease of entropy for the outside region of the black hole is
compensated by the increase of the Bekenstein-Hawking entropy of the
black hole caused by the work done by the observer at the other end of
the string located at infinity, as Unruh and Wald~\cite{UW1,UW2} showed by
taking into account the buoyancy force of the thermal atmosphere of
the quantum fields surrounding the black hole. In the
present work, we pursue this direction of research by elucidating
the role of the entanglement between the states of the
inside and the outside regions
of a black hole, and by applying a quantum information theoretical
approach \cite{PERE} to a similar gedanken experiment in the black hole
spacetime.
This is also motivated by the so called `holographic principle' approach
\cite{TH,FS,BO},
according to which the information $S_V$ contained within any region $V$
is upper bounded by the area $A_B$ of its boundary ($S_V\leq A_B/4$).

We shall start with an elementary introduction of quantum information
entropy, and then go over to the generalized second law of black hole
thermodynamics.

\section{Entanglement entropy and role of the detector in the second law}

Let us start by discussing the ordinary second law by using quantum
information theory \cite{PERE}.
This will elucidate the role of the entanglement of the states for the
inside and
outside region of a black hole, and especially the important role of the
detector in a measurement process.

Let the initial state be
\begin{eqnarray}
|\psi>=\sum_{n}\sqrt{c_{n}}|n>_{B}|n>_{A}|\Phi_{0}>~,
\label{psi1}
\end{eqnarray}
in a Schmidt decomposition form.
Here $|n>_{A}$ is Alice's state and $|n>_{B}$ is Bob's state, which
are entangled.
We have also introduced a detector state, which is initially in the ground
state $|\Phi_{0}>$.
Note that, without loss of generality,
we can assume that the $c's$ are ordered, i.e. that
$c_{0}\geq c_{1}\geq \ldots \geq 0~$,
and that the detector state is either $|\Phi_{0}>$ or $|\Phi_{1}>$.

Suppose further that we are completely ignorant about Bob's states
so that we have to trace over them
to obtain a density matrix
for Alice's and the detector states, i.e.
\begin{eqnarray}
\rho=Tr_{B}|\psi><\psi|=\sum_{n}c_{n}|n>_{A}|\Phi_{0}><\Phi_{0}|<n|_A~,
\end{eqnarray}
and the corresponding von Neumann entropy reads
\begin{eqnarray}
S=-Tr(\rho \log \rho)=-\sum_{n}c_{n}\log c_{n}~,
\end{eqnarray}
which is also called the entanglement entropy.

Now let us switch on the apparatus so that Alice and the detector
states undergo the unitary evolution $U^{\alpha}_{nm}$ ($\alpha=0, 1$), i.e.
\begin{eqnarray}
|\psi> &\rightarrow& |\psi'> =\sum_{n}\sqrt{c_{n}}|n>_{B}
\sum_{m,\alpha}U^{\alpha}_{nm}|m>_{A}|\Phi_{\alpha}>~,
\end{eqnarray}
and then switch off the apparatus again.
The measurement process then involves a separation of the wave packet into
macroscopically distant ones corresponding to the {\lq}{\lq}collapse of
the wave function{\rq}{\rq} (e.g., imagine an experiment \`{a} la
Stern-Gerlach) for different values of $\alpha$.
For each $\alpha$, the density matrix
and the weighted mean entropy ${\bar S}'$ are given by
\begin{eqnarray}
\rho'^{\alpha}
&=&\sum_{n,m,m'}\frac{c_{n}}{K_{\alpha}}U^{\alpha}_{nm} U^{*\alpha}_{nm'}
|m>_{A}|\Phi_{\alpha}><\Phi_{\alpha}|<m'|_{A}~,
\nonumber \\
{\bar S}'&=&\sum_{\alpha}K_{\alpha}S'^{\alpha}~~;
S'^{\alpha}=-Tr(\rho'^{\alpha}\log\rho'^{\alpha})~,
\label{roprime}
\end{eqnarray}
where $K_{\alpha}=\sum_{m,n}c_{n}|U^{\alpha}_{nm}|^{2}$ is the probability
to get the state $|\Phi_{\alpha}>$ by measurement, and
$\sum_{\alpha}K_{\alpha}=1$.
Moreover, we can show from the concavity of the entropy that
${\bar S}'\leq
S(\sum_{\alpha}K_{\alpha}\rho'^{\alpha}) =
S(\rho)~$.
Therefore the average entropy decreases due to the measurement process.

We then do a local unitary transformation
$T^{\alpha}$ to Bob's state $|n>_{B}$ knowing the outcome $\alpha$ of the
experiment via classical communication,  so that the resultant state is of
the Schmidt form,
\begin{eqnarray}
|{\psi}''^{\alpha}>
=\sum_{m}{1\over \sqrt{K_{\alpha}}}W^{\alpha}_{m}|m>_{B}|m>_{A}|\Phi_{\alpha}>.
\label{schmidt}
\end{eqnarray}
Here the unitary matrix $T^{\alpha}$ is arranged so that
$\sum_{n}\sqrt{c_{n}}T^{\alpha}_{nl}U^{\alpha}_{nm}=W^{\alpha}_{m}
\delta_{l,m}$ holds.
For each $\alpha$ the density matrix is
$\rho''^{\alpha}
=[\sum_{m}|W^{\alpha}_{m}|^{2}|m>_{A}<m|_{A}]/K_{\alpha}$
and  the corresponding conditional entropy is
$S''^{\alpha}
=\log
K_{\alpha}-[\sum_{m}|W^{\alpha}_{m}|^2\,log\,|W^{\alpha}_{m}|^2]/K_{\alpha}$.
Then the total entropy sum of the detector entropy
\begin{eqnarray}
S^{d}=-\sum_{\alpha}K_{\alpha}log K_{\alpha}~,
\end{eqnarray}
and of the average entropy, ${\bar S}''\equiv \sum_{\alpha}K_{\alpha}
S''^{\alpha}~$ (${\bar S}''=\bar{S}'$),
can be easily shown to increase,
\begin{eqnarray}
{\bar S}''+S^{d}\geq S~.
\end{eqnarray}

Therefore, the apparent violation of the ordinary second law due to the
decrease in the average matter entropy is avoided by properly taking into
account the role of the detector entropy.

One might note that our model (as for the decrease of the average
entropy after measurement, eqs. (2-6)) is a particular realization of the
general framework of Ref. \cite{NI}, where it is shown that by local
operations which
consist of unitary transformations of Alice's and the detector states, and
via a POVM process, one can have the following transition
$\sum_{n}\sqrt{c_{n}}|n>_{B}|n>_{A}
\rightarrow \sum_{n}\sqrt{c_{n}'}|n>_{B}|n>_{A}~$,
with $c'_{0}\geq c_{0}~,~c'_{0}+c'_{1}\geq c_{0}+c_{1}~,~\ldots~, c'_{0}+
c'_{1}+\cdots =c_{0}+c_{1}+\cdots =1$, i.e. where the $\{c'\}$'s
{\lq}{\lq}majorize{\rq}{\rq} the $\{c\}$'s.
In our case,
$c'_{n}=|W^{\alpha}_{n}|^2\equiv \sum_{m}c_{m}|U^{\alpha}_{mn}|^2$,
and the majorization implies that
$~-\sum_{n}c'_{n}\log c'_{n}\leq -\sum_{n}c_{n}\log c_{n}~$
(from the concavity of the entropy).

\section{Bekenstein's gedanken experiment reexamined}

As it is by now well-known, the radiation state of an eternal black hole
can be described (in a semiclassical approximation for the background
geometry) by the Hartle-Hawking state :
\begin{eqnarray}
|\psi_{HH}>=\prod_{\omega}\sum_{n}\sqrt{c_{n}}|n>_{B}|n>_{A}~,
\end{eqnarray}
where $c_{n}=e^{-\frac{n\omega}{T_{BH}}}/Z$, $T_{BH}$ is the Hawking
temperature of the black hole, $Z$ is the partition function,
$|n>_{B}$ is the $n$-particle state for the region inside the event horizon
and $|n>_{A}$ is that for the region outside the event
horizon.
Hereafter, to simplify the presentation, we will consider only a particular
mode of the particle states with angular frequency $\omega$.
The state $|\psi_{HH}>$ is an entangled state of Alice's and Bob's states
(Alice and Bob can see only $|n_A>$ and $|n_B>$, respectively), which
is naturally prepared in general relativity.

Imagine we do a local operation by observing Alice's states. Tracing
over Bob's states, we see that the entanglement entropy decreases as
we have shown in the previous section. Of course, this does not necessarily
imply the violation of the second law, since by taking into account the
detector entropy the total entropy will still increase as we have seen before.

However, what happens if the detector carrying the entropy acquired by
the measurement is dropped into the black hole ? It seems that the
loss in radiation entropy caused by the local operation has no way to be
compensated by the detector entropy, which is now gone into the black hole.
This is a quantum analogue of the {\lq}{\lq}classical{\rq}{\rq} gedanken
experiment proposed by Bekenstein \cite{BE1,BE2} and discussed by Unruh
and Wald~\cite{UW1,UW2}.

Before discussing our {\lq}{\lq}quantum{\rq}{\rq} gedanken experiment,
we reexamine Bekenstein's {\lq}{\lq}classical{\rq}{\rq} gedanken
experiment by exploiting a fundamental inequality of ordinary thermodynamics
between the work done on a system and the change of the total
Helmoltz free energy.
Suppose that the black hole has reached thermal equilibrium with the
surrounding radiation, the whole system being enclosed in a large
cavity whose boundary is located far from the horizon so that its
temperature equals the Hawking temperature $T_{BH}$.
Then, the local temperature of the thermal radiation is given by
Tolman's law as $\tilde{T}=T_{BH}/\chi$, where $\chi$ is the local
redshift factor.
In the gedanken experiment \`{a} la Bekenstein, the box containing some
energy $E^{b}$ and entropy $S^{b}$ is initially located outside the
cavity (A in Fig.~1-a),
then slowly lowered down by means of a string towards the black hole
(B in Fig.~1-b)
and finally dropped into the black hole from some point near the horizon
(Fig.~1-c).

In this process, the change in the total entropy is
\begin{eqnarray}
\Delta S^{tot} &=& \frac{\Delta M}{T_{BH}} + \Delta S^{m}~,
\label{eqn:total-entropy-change1}
\end{eqnarray}
where $\Delta M$ and $\Delta S^{m}=-S^{b}$ are the change in the black
hole mass and that in matter entropy, respectively. Furthermore, the energy
conservation law implies that
\begin{eqnarray}
\Delta M &=& E^{b} + \Delta W~, \label{eqn:ADM-mass}
\end{eqnarray}
where $E^{b}$ is the initial energy of the box and $\Delta W$ is the
work done by the agent at infinity in this process. Note that, since
$\Delta M$ is the change in the ADM mass within the cavity, $\Delta W$
is the work done by the agent at infinity to the whole system of the
black hole and the thermal atmosphere.

Noting that the value of the temperature at the boundary of the
cavity is fixed during this process,
we can evaluate the work $\Delta W$ done to the system at infinity
during this isothermal process by using the ordinary thermodynamics
as \cite{LL}
\begin{eqnarray}
\Delta W \ge F_{f}- F_{i}~,
\label{eqn:work}
\end{eqnarray}
 where $F_{f}$ ($F_{i}$) is the Helmholtz free energy of the
final (initial) stage observed from infinity\footnote{
Similar arguments can be easily generalized to the case of a charged,
rotating (quasistationary) black hole background.}.
The equality in Eq.~(\ref{eqn:work}) holds for a quasistatic process.
Assuming that the lowering process does not disturb the bulk of the
thermal radiation and neglecting the terms which cancel after the
subtraction, each term in the right hand side of the
inequality~(\ref{eqn:work}) can be written as
\begin{eqnarray}
F_{f}&=&(E^{b}_{\ast}-T^{b}_{\ast}S^{b}_{\ast})\chi_{\ast}~;
\nonumber \\
F_{i}&=&(E^{b}-T^{b}_{\infty}S^{b})
+(E^{r}_{\ast}-\tilde{T}_{\ast}S^{r}_{\ast})
\chi_{\ast}~,
\label{eqn:free-energy2}
\end{eqnarray}
where the quantities $E^{r}$ and $S^{r}$ are the energy and entropy of
the displaced thermal radiation,
the index $\ast$ means that the quantities are evaluated at the dropping
point and $\chi$ denotes the red-shift factor.
Assuming that
the content of the box is completely shielded from the surrounding radiation
by impenetrable walls, we can see that for the box the temperature
$T^{b}_{\ast}\chi_{\ast}=T^{b}_{\infty}$,
and the energy and entropy remain constant,
$E^{b}_{\ast}=E^{b}$, $S^{b}_{\ast}=S^{b}$.

Therefore,
putting Eqs.~(\ref{eqn:total-entropy-change1})-(\ref{eqn:free-energy2})
together, we get
\begin{eqnarray}
\Delta S^{tot}
&\ge& \left(S^{r}_{\ast}
-{E^{r}_{\ast}\over T_{BH}}\chi_{\ast}\right) -\left(S^{b}
-{E^{b}\over T_{BH}}\chi_{\ast}\right)~,
\label{eqn:total-entropy-change2}
\end{eqnarray}
where $E^{r}_{\ast}$, $S^{r}_{\ast}$ and $\chi_{\ast}$ are functions of
the distance $r$ of the box from the horizon. Eq.
(\ref{eqn:total-entropy-change2}) corresponds to Eqs.~(16) and (A12) of
Ref.~\cite{UW2}, although we have not used model dependent arguments like that
on the buoyancy force.
The minimum of the function $\Delta S^{tot}(r)$ can be found by minimizing the
change in the ADM mass $\Delta M$, i.e.
$d[\Delta S^{tot}(r)]/dr
=[(E^{b}-E^{r}_{\ast})/T_{BH}](d\chi_{\ast}/dr)=0~$,
where we have used the first law of thermodynamics
for the surrounding radiation. Using the terminology of Ref. \cite{UW2},
one can see
that the minimum is realized when the box is dropped into the black hole
from the floating point $E^{b}=E^{r}(r_{fp})$ where the gravity and
buoyancy force on the box balance. Therefore, it is sufficient for the
estimation of the lower bound on $\Delta S^{tot}$ to consider the case
when
we cut the string at the floating point $r_{fp}$ to let the box freely
fall into the black hole:
\begin{eqnarray}
\Delta S^{tot}
\ge \left(S^{r}_{\ast}-S^{b}\right) |_{r=r_{fp}} \ge 0,
\label{eqn:lower-bound}
\end{eqnarray}
where the last inequality follows from the fact that the maximum entropy is
achieved by the thermal distribution for fixed values of volume and energy.

Therefore, by using only the physical properties of ordinary matter, we can
show that the total entropy~(\ref{eqn:total-entropy-change1}) always
increases in this gedanken experiment within the first order approximation,
i.e., when the changes in black hole mass and matter entropy are assumed to be
small compared with the values of black hole mass and entropy themselves, and
we can neglect the backreaction effects.

We would like to emphasize that the generalized second law makes sense only
when we consider the entropy of the outside region and can be shown through
the ordinary second law of thermodynamics, assumed to hold locally in the
outside region.

\section{The generalized second law and quantum information}

Now consider the {\lq}{\lq}quantum{\rq}{\rq} gedanken experiment which might
seem to cause the violation of the generalized second law at first sight.
Let us imagine a situation in which we perform a quantum measurement
that splits
the wave packet, initially prepared on the hypersurface $\Sigma$ as in Eq.
\rf{psi1}\footnote{The generalized second law can be shown to hold even
if the initial state of the detector
on $\Sigma$ is mixed.}, according to the
detector states $|\Phi_{0}>$ and $|\Phi_{1}>$ on the hypersurface $\Sigma'$.
On the final hypersurface $\Sigma''$, only one of the packets (e.g.,
$|\Phi_{0}>$) remains at the end of the string, while the other one (e.g.,
$|\Phi_{1}>$) is freely falling into the black hole (see Fig.~2).
On the hypersurface $\Sigma''$, the reduced density matrix $\rho''^{0}$
obtained by tracing over Bob's states is given by eq. \rf{roprime}
with $\alpha=0$.

By using Nielsen's inequality,
we have an apparent violation of the second law
\begin{eqnarray}
S''^{0}\equiv -Tr_{A}(\rho''^{0}\log \rho''^{0})
\leq S=-\sum_{n}c_{n}\log c_{n}~,
\end{eqnarray}
which is caused by the local operation involving the measurement.

However, the change in total entropy is
\begin{eqnarray}
\Delta S^{tot}={\Delta W\over T_{BH}}+\Delta S^{m}
\geq\frac{F_{II}-F_{I}}{T_{BH}}+{\bar S}''-S^r~,
\end{eqnarray}
where $F_{II}$ ($F_{I}$) is the free energy of the detector and
radiation after (before) the local operation. These quantities are evaluated
in the region $C$ in Fig.~2, and we obtain
\begin{eqnarray}
F_{II}&=&[K_{0}({E''}_{0}-\tilde{T}S''^{0}) +K_1({E''}_1
-\tilde{T}{S''}^{1})]\chi~,
\nonumber\\ 
F_{I}&=&(E-\tilde{T}S^{r})\chi~,
\label{deltaf}
\end{eqnarray}
where $\chi$ is the redshift factor in the small region $C$,
${E''}_{\alpha}$ the energy of the detector and radiation for a given
$\alpha$, and $\tilde{T}\chi=T_{BH}$ as before.

We are now close to the conclusion, since
\begin{eqnarray}
\Delta S^{tot}\geq 0~,
\end{eqnarray}
due to the energy conservation,  $K_{0}{E''}_{0}+K_{1}{E''}_{1}=E$.

\section{Summary}

A quantum version of Bekenstein's  ``classical'' gedanken experiment has
been proposed.
We performed a series of local operations which consist of
certain unitary transformations for the outside and the detector states
and a local measurement in the region outside the black hole. 
In general, this causes a decrease of the entanglement entropy by Nielsen's 
inequality.
However, as far as the detector remains outside, the total entropy including 
the detector one always increases.
When the detector is conditionally dropped into the black hole depending on
the experimental outcome, 
the decrease of the matter entropy, $\Delta S^{m}=-S^{b}$, caused by
dropping the box is more than compensated by the increase
of the black hole entropy, $\Delta S^{BH}= \Delta M / T_{BH}$,
which ultimately arises from the work done by the measurement
via the first law.
Therefore, the generalized second law holds.

We here comment briefly on the work by Frolov and Page~\cite{FP}.
Although they proved the generalized second law for eternal black holes
by comparing the initial
vacuum state before the collapse of a star and the final entangled state,
their situation is quite different from our gedanken experiment
(e.g., nothing equivalent to the detector is thrown into the black hole).
Finally, we would like to make the following remark.
For the change of the total entropy in the gedanken experiment of Section III,
Unruh and
Wald have derived the same expression as Eq.~(\ref{eqn:total-entropy-change2})
by using the buoyancy force exerted on the box by the black hole atmosphere.
So far, several
discussions about the practical effects of this force on the energetics of
the process have been made ~\cite{W}.
In particular, Bekenstein recently argued~\cite{BE3} that the computation of
the Unruh-Wald buoyancy force using the stress-energy tensor is invalid for
the case in which the size of the box is smaller than the typical wavelength
of the accelerated radiation,
and that his entropy bound $S^b/E^b\leq 2\pi R$ ($R$ being the
characteristical size of the box) is
necessary in order for the generalized second law to be valid.
In our derivation,
we have not used model dependent arguments like the one on the buoyancy
force, and this is in the spirit
that the generalized second law should be a fundamental law of nature.

\noindent {\Large \bf Acknowledgements}

\bigskip
A.H.was partially supported by the Ministry of Education,
Science, Sports and Culture of Japan, under grant n. 09640341.
A.C. was supported by JISTEC, under a STA grant n. 199016; 
he also thanks the cosmology group at TITECH and the photonic 
group at CRL for the kind hospitality during this work.

\begin{figure}
\begin{center}
\leavevmode
\epsfysize=8.0cm
\epsfbox{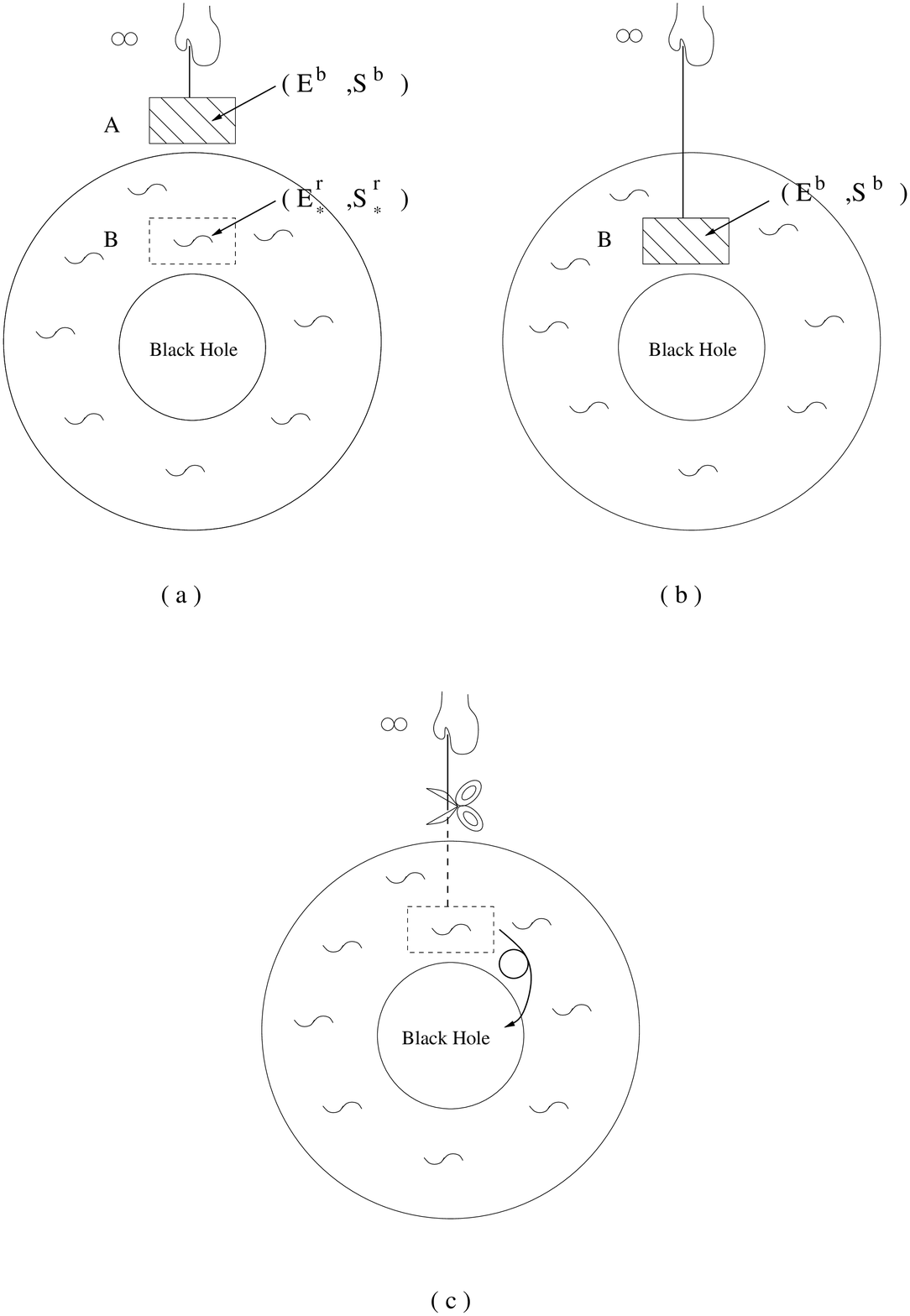}
\end{center}
\caption[]
{The box is initially located outside the cavity and then lowered down 
towards the region near the horizon.
Finally, we cut the string and the box is thrown into the black hole.}
\label{fig:total-system}
\end{figure}
\begin{figure}
\begin{center}
\leavevmode
\epsfysize=4.0cm
\epsfbox{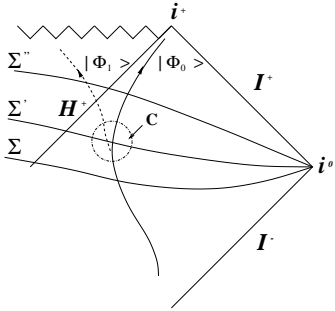}
\end{center}
\caption[]
{We prepare the wave packet on the hypersurface $\Sigma$,
and then perform a quantum measurement which splits the wave packet
into macroscopically distant ones, 
according to the detector states $|\Phi_{0}>$ and $|\Phi_{1}>$ 
on the hypersurface $\Sigma'$. 
On the final hypersurface $\Sigma''$ only one of the packets, 
$|\Phi_{0}>$, remains at the end of the string, while the other, 
$|\Phi_{1}>$, is freely falling into the black hole.}
\label{fig:gedanken-experiment}
\end{figure}

\end{document}